\documentclass[aps,showpacs,manuscript,12pt]{revtex4}
\usepackage{amssymb}
\usepackage{amsmath}
\usepackage{graphicx}

\begin{document}

\title{Nonlinear Debye screening in strongly-coupled plasmas}
\author{D. Sarmah$^{1,2,3}$, M. Tessarotto$^{1,3}$ and M. Salimullah$^{4,5}$}
\affiliation{$^{1}$Department of Mathematics and Informatics,
University of Trieste, Trieste, Italy \\
$^{2}$ International Center for Theoretical Physics, ICPT/TRIL
Program, Trieste, Italy \\
$^{3}$ Consortium for Magnetofluid Dynamics, Trieste, Italy\\
$^{4}$ Department of Physics, Jahangirnagar University, Savar,
Dhaka, Bangladesh\\
$^{5}$ Salam Chair in Physics and Department of Physics, GC
University, Lahore, Pakistan }

\begin{abstract}
An ubiquitous property of plasmas is the so-called Debye shielding
of the electrostatic potential. Important aspects of Debye
screening concern, in particular, the investigation of non-linear
charge screening effects taking place in strongly-coupled plasmas,
that imply a reduction of the effective charge characterizing the
Debye-H\"{u}ckel potential. These effects are particularly
relevant in dusty plasmas which are characterized by high-Z
particles. The investigation of the effective interactions of
these particles has attracted interest in recent years especially
for numerical simulations. In this work we intend to analyze the
consistency of the traditional mathematical model for the Debye
screening. In particular, we intend to prove that the 3D Poisson
equation involved in the DH model does not admit strong solutions.
For this purpose a modified model is proposed which takes into
account the effect of local plasma sheath (i.e., the local domain
near test particles where the plasma must be considered discrete).
Basic consequences of the model are discussed, which concern the
asymptotic properties of the solutions determined both for weakly
and strongly coupled plasmas. As an application the charge
screening effect in strongly coupled plasmas is investigated and
an explicit expression of the effective charge for the asymptotic
DH potential is determined.
\end{abstract}

\date{\today }
\maketitle

\section{Introduction}

A basic aspect of plasma physics is the so-called Debye shielding
of the electrostatic potential. \ This consists in the property of
plasmas (or electrolytes \cite{Debye 1923}), either quasi-neutral
or non-neutral, to shield the electrostatic field produced by
charged particles (to be also denoted as test particles) immersed
in the same system. This result has fundamental consequences on
plasma phenomenology, since it actually limits the range of static
Coulomb interactions inside the Debye sphere, i.e., at a distance
$\rho \leq \lambda _{D}$ from the test particle$,$ $\lambda _{D}$
being the Debye length. As usual here $\lambda _{D}\equiv \left(
\sum\limits_{s}\lambda _{Ds}^{-1}\right) ^{-1},$ where the sum is
carried out on all plasma species and $\lambda
_{Ds}=\sqrt{\frac{T_{s}}{4\pi
Z_{s}^{2}e^{2}N_{os}}},$ $T_{s}$ and $N_{os}$ being respectively the $s-$%
species temperature and number density (the latter defined in the
absence of test particles). In fact, when both the particles and
the plasma are assumed non-relativistic, small-amplitude,
stationary (or slowly time- and space-varying), electrostatic
perturbations generated by isolated test particles, result
effectively shielded in the external domain, i.e., at distances
larger than the Debye length $\lambda _{D}$. The renewed interest
in this problem \cite{Robbins-1988,Dupont 1996,Allayrov
1998,Bystrenko 1999,Tsytovich 2001,Bystrenko
2003,Tessarotto2006,Tessarotto2006b} is particularly related to
dusty plasmas or colloidal suspensions which are characterized by
the presence of a large fraction of highly charged
particles (grains), i.e., having an electric charge $Z_{d}e$ with $%
\left\vert Z_{d}\right\vert \gg 1$.

In this work we intend to analyze the consistency of the
traditional mathematical model for the so-called \emph{Debye
screening problem} (DSP) originally formulated by Debye and
H\"{u}ckel (\emph{DH model }\cite{Debye 1923}). In particular, we
intend to prove that the 3D Poisson equation involved in the DH
model does not admit physically acceptable solutions, i.e.,
solutions which are provided by ordinary functions and are at
least continuous in the domain of existence, i.e., are so-called
classical (or strong) solutions. For this purpose a modified model
is proposed which takes into account the effect of local plasma
sheath (i.e., the local domain near test particles where the
plasma must be considered discrete). Basic consequences of the
model are discussed, which concern the asymptotic properties of
the solutions determined both for weakly and strongly-coupled
plasmas.

Despite previous attempts to construct approximate or exact
solutions to the DH model \cite{Perez 1998,Chang Lin 2000,Martin
1994}, the related mathematical model appears incomplete and can
be shown to be physically unacceptable, due to the neglect of the
local plasma sheath. In fact, it is obvious that sufficiently
close to the point-particle the so-called
weak-field approximation is violated making the DH model invalid \cite%
{Tessarotto2006}. In the past \cite{Lampert 1980} it was pointed
out that in such a case the test particle does not produce any
electric field, but only complete charge neutralization by the
plasma, thus producing a Debye length which effectively vanishes.
Other objections concerned the asserted indeterminacy of the
solution for $x=0$ due to its divergence in the same point
\cite{Lo Surdo 1994}. These issues were later addressed in a more
general context \cite{Garrett 1988}, including the 2D case where
complete neutralization cannot be achieved. To recover the correct
physical picture the effect of local plasma sheath must be
included. Nevertheless, for suitably dense plasmas or in the case
of plasma species characterized by very high electric charges
(high-Z), such as dusty plasmas, the weak-field approximation may
be locally violated. This circumstance, when the effect of finite
local plasma sheath is included, occurs if the normalized
electrostatic potential $\widehat{\Phi }(\rho )$ results of order
unit or larger on the boundary of the plasma sheath (produced by
at least one of the $s$ plasma species), namely for $\rho =\rho
_{os}.$

It is well-known that, in general, the Debye effect occurs
provided suitable physical assumptions are introduced. In
particular, the plasma must be assumed appropriately close to
kinetic Maxwellian equilibrium, in which each particle species is
described by a Maxwellian kinetic distribution function carrying
finite fluid fields [defined respectively by the number density,
temperature and flow velocity $\left( N,T,\mathbf{V}\right) $]. In
the absence of test particles these fluid fields must be assumed
slowly varying
in a suitable sense, or constant, both with respect to position ($\mathbf{r}$%
) and time ($t$). \ In this regard it is important to remark that
the appropriate treatment of the plasma sheath surrounding each
test particle is essential also for the validity of the
mathematical model for the Debye screening problem, i.e., for the
existence of classical solutions of the
Debye screening problem, which do not exist when letting $x_{o}=0$ \cite%
{Tessarotto2006}.

Another significant aspect concerns the issue of the absorption of
plasma particles by the test particle, which effectively modifies
the local charge
density of the background plasma species \cite%
{Bernstein1959,Alpert1965,Allen1992}. Since the particle capture
mechanism is a manifestly charge-dependent and velocity-dependent
phenomenon (in particular it depends on the angular momentum of
the incoming particle), it is obvious that in principle it can
produce deviations from local Maxwellian equilibrium
\cite{Allen1999,Tsytovich2005}. However, this phenomenon is
expected to become relevant only if the radii of the test particle
and of the surrounding plasma sheath ($\rho _{p}$ and $\rho _{o}$)
are comparable, i.e., $\rho _{p}/\rho _{o}\sim 1$. Instead, is
results negligible when $\rho _{p}/\rho _{o}\ll 1$. Since dusty
and colloidal
plasmas are characterized by typical grain size $\rho _{p}$ smaller than $%
10^{-4}-10^{-5}$ $cm$ and radius of plasma sheath $\rho _{o}$ of
the order of $10^{-2}-10^{-3}$ $cm$, these effects will be
considered negligible.

Goal of this work is the analysis of DSP and the definition of a
suitably modified mathematical model to take into account the
effect of local plasma sheaths in quasi-neutral plasmas. In
particular, in Sec. 2 a modified Debye screening problem (modified
DSP) is presented. In Sec. 3 the basic mathematical result is
presented which concerns the non-existence of classical solutions
of DSP. The proof is reached by noting that DSP can be obtained as
limit problem obtained from the modified DSP. Basic feature of the
approach is the representation of the Poisson equation in integral
form. This permits to analyze the asymptotic properties of the
solutions of the modified problem in the limit $x_{os}\rightarrow
0^{+}.$ It is found, that the limit solution of the modified DSP
for $x_{os}\rightarrow 0^{+}$ is a distribution which vanishes
identically for all $\rho >0$ and is discontinuous in $\rho =0.$
Hence, the limit function $\widehat{\Phi }(x)$ is not an strong
solution of the DSP equation. This is therefore a characteristic
property of the DH model. In particular, as a basic consequence,
the effective charge of the DH asymptotic solution $c$ vanishes
identically in such a limit and results independent of the charge
of the test particle.

\section{The modified DSP}

The traditional formulation of the DSP, based on the
Debye-H\"{u}ckel model \cite{Debye 1923} regards the test
particles as point-like and having a spherically-symmetric charge
distribution while ignoring the effect of local plasma sheath.
This implies, from the physical standpoint, to neglect the
discrete nature of the plasma. Here we shall consider a modified
modified
Debye-H\"{u}ckel model, based on the introduction of the notion of \textit{%
local plasma sheath }\cite{Tessarotto2006}. In the sequel we shall
consider for simplicity of notation the case of a two
species-plasma, formed by electrons and Hydrogen ions, having an
unique plasma sheath. Thus, we shall assume that the test particle
is represented by a spherically symmetric charge of radius $\rho
_{p}.$ For a particle in which $\rho _{p}<\rho _{o}$ the plasma
sheath is represented by the spherical shell centered at the
position (center) of the test particle for which $\rho _{p}\leq
\rho <\rho _{o}$, in which the plasma charge density (except for
the presence of the test particle) results negligible. In the
sequel we can also let in particular $\rho _{p}=0$ (point-like
test particle) or $\rho _{p}=\rho _{o}$ (finite-size test
particle). The customary DH model is thus recovered letting $\rho
_{p}=0$ and taking the limit $\rho _{o}\rightarrow 0$ (or in
dimensionless variables, requiring $x_{p}\equiv \rho _{p}/\lambda
_{D}=0$
and $x_{o}\equiv \rho _{o}/\lambda _{D}$\ $\rightarrow 0$). Denoting $%
\widehat{\Phi }_{x_{o}}(x)$ the solution of the Poisson equation,
here we intend to determine its asymptotic properties in the limit
$x_{o}\rightarrow 0^{+},$ while\ also letting $x_{p}=0$ (see
Lemma). As a consequence and in agreement with \cite{Lampert
1980,Garrett 1988}, in such a case it follows
that the limit function $\lim_{x_{o}\rightarrow 0^{+}}\widehat{\Phi }%
_{x_{o}}(x)\equiv \widehat{\Phi }(x)$ vanishes identically for $x>0,$ i.e., $%
\lim_{x_{o}\rightarrow 0^{+}}\widehat{\Phi }_{x_{o}}(x)=0.$In
addition, in the same set we intend to prove the identity
\begin{equation}
\beta -\lim_{x_{o}\rightarrow 0^{+}}\int_{x_{o}}^{x}dx^{\prime
}x^{\prime 2}\sinh \widehat{\Phi }_{x_{o}}(x^{\prime
})\widehat{\Theta }(x^{\prime }-x_{o})=0,  \label{Eq.3-3}
\end{equation}%
where $\widehat{\Theta }(x-x_{o})$ us the (weak) Heaviside
function. In detail the relevant equations valid in each subdomain
for the normalized electrostatic potential $\widehat{\Phi
}_{x_{o}}(x\mathbf{)}$ are as follows. In the internal domain
$0\leq x<x_{p}$ the electrostatic potential
is assumed constant, namely $\widehat{\Phi }_{x_{o}}(x\mathbf{)=}\widehat{%
\Phi }_{x_{o}}(x_{p}\mathbf{)}$. In the plasma sheath $x_{p}\leq x<x_{p},$ $%
\widehat{\Phi }_{x_{o}}(x)$ satisfies the customary Poisson
equation in the presence of the charge density produced by a
finite-size spherically-symmetric charge
\begin{equation}
\nabla _{x}^{2}\widehat{\Phi }_{x_{o}}=-\frac{\beta }{x^{2}}\delta
(x-x_{p}). \label{DSP-modified -b}
\end{equation}%
Finally, in the external domain $x>x_{o}$ there holds the Poisson
equation in the presence of the plasma charge density:
\begin{equation}
_{x}^{2}\widehat{\Phi }_{x_{o}}=\widehat{\Theta }(x-x_{o})\sinh \widehat{%
\Phi }_{x_{o}}.  \label{DSP modified}
\end{equation}%
The boundary conditions, imposed respectively at infinity and at
the boundary of the plasma sheath, are specified as follows
\begin{align}
\lim_{x\rightarrow \infty }\widehat{\Phi }_{x_{o}}(x\mathbf{)}&
=0,
\label{BC-1 b} \\
\left. x^{2}\frac{d}{dx}\widehat{\Phi
}_{x_{o}}(x\mathbf{)}\right\vert _{x=x_{o}}& \mathbf{=-}\beta .
\label{BC-2 b}
\end{align}%
We notice that, if $x_{p}<x_{o}$ (for example, $x_{p}=0$)$,$ $\widehat{\Phi }%
_{x_{o}}(x\mathbf{)}$ results by assumption at least of class $C^{(1)}(%
%TCIMACRO{\U{211d} }%
%BeginExpansion
\mathbb{R}
%EndExpansion
_{\{x_{p}\}}),$ where $%
%TCIMACRO{\U{211d} }%
%BeginExpansion
\mathbb{R}
%EndExpansion
_{\left\{ x_{p}\right\} }\equiv \left] x_{p},\infty \right[ $ . Here $%
x_{o},\beta $ are both assumed constant and strictly positive real
numbers. The problem defined by (\ref{DSP
modified}),(\ref{DSP-modified -b}), together with the boundary
conditions (\ref{BC-1 b}),(\ref{BC-2 b}), will be
here denoted as \emph{modified DSP. }\ From the physical standpoint Eqs.(\ref%
{DSP modified}),(\ref{DSP-modified -b}) may be viewed as the
Poisson equation for a spherical ideally conducting charge, or for
a point particle
in the presence of a plasma sheath, of radius $r_{o}$ (i.e., $%
x_{o}=r_{o}/\lambda _{D}$ in non-dimensional variables) which is
in electrostatic equilibrium and is immersed in a spatially
uniform quasi-neutral and Maxwellian plasma. As for the previous
DSP equation, it
follows that, for solutions satisfying the boundary conditions (\ref{BC-1 b}%
),(\ref{BC-2 b}), in the domain $x\in
%TCIMACRO{\U{211d} }%
%BeginExpansion
\mathbb{R}
%EndExpansion
_{\left\{ x_{o}\right\} }$ Eq.(\ref{DSP modified}) can be cast in
the
integral form%
\begin{align}
\widehat{\Phi }_{x_{o}}(x)& \mathbf{=}\frac{\beta \widehat{\Theta }(x-x_{o})%
}{x}-  \label{integral form - DSP modified} \\
& -\left[ \frac{1}{x}\int_{x_{o}}^{x}dx^{\prime }x^{\prime
2}+\int_{x}^{\infty }dx^{\prime }x^{\prime }\right] \sinh \widehat{\Phi }%
_{x_{o}}(x^{\prime })\widehat{\Theta }(x^{\prime }-x_{o}).  \notag
\end{align}%
In particular, thanks to continuity at $x=x_{o}$ of $\widehat{\Phi }%
_{x_{o}}(x),$ one obtains the constraint
\begin{equation}
\widehat{\Phi }_{x_{o}}(x_{o})=\Gamma -\int_{x_{o}}^{\infty
}dx^{\prime
}x^{\prime }\sinh \widehat{\Phi }_{x_{o}}(x^{\prime })\widehat{\Theta }%
(x^{\prime }-x_{o}),  \label{constraint for Fi}
\end{equation}%
with $\Gamma \equiv \frac{\beta }{x_{o}}$ denoting the Coulomb
coupling
parameter. It is immediate to establish, the existence and uniqueness of $%
\widehat{\Phi }_{x_{o}}(x)$ in the functional \ class
$\widehat{C}^{(\infty
)}(%
%TCIMACRO{\U{211d} }%
%BeginExpansion
\mathbb{R}
%EndExpansion
_{\left\{ x_{o}\right\} }),$ together with its continuous
dependence on
initial data, in particular the continuity with respect to the parameter $%
x_{o}\in _{\left\{ 0\right\} }.$ Moreover, assuming that the
weak-fields approximation applies (this condition is manifestly
fulfilled identically in the weak-coupling ordering,$\ 0<\Gamma
\sim O(\varepsilon )\ll 1),$ and is satisfied at least for $x\gg
1$ suitably large$,$ it is immediate to prove that in this subset
an asymptotic\ solution of the modified DSP is provided by the
\emph{external asymptotic solution}
\begin{equation}
\widehat{\Phi }_{x_{o}}(x)\cong \widehat{\Phi
}_{x_{o}}^{(ext)}(x)\equiv \frac{c}{x}e^{-x+x_{o}}.
\label{external asymptotic solutionl}
\end{equation}%
Here denoted as of the modified DSP and $c=c(x_{o},\Gamma ).$is the \emph{%
effective dimensionless charge.} Hence, $\widehat{\Phi
}_{x_{o}}^{(ext)}(x)$ reduces formally to the so-called DH
potential when\ $x\gg x_{o}.$ In the weak-coupling ordering it
follows $c(x_{o},\Gamma )=\frac{\beta }{1+x_{o}},$ while for
strongly-couple plasmas a lower value is expected. Furthermore, it
is obvious that the limit function $\lim_{x_{o}\rightarrow 0^{+}}\widehat{%
\Phi }_{x_{o}}(x)$ coincides with the solution of DSP, i.e.,%
\begin{equation}
\lim_{x_{o}\rightarrow 0^{+}}\widehat{\Phi }_{x_{o}}(x)=\Phi (x).
\label{limit function and DSP}
\end{equation}

\section{Non-existence of classical solutions of DSP}

Let us now analyze, which consequences can be obtained for the Debye-H\"{u}%
ckel problem, formally obtained by letting $x_{o}=0$ in the
previous equations [in particular Eq.(\ref{integral form - DSP
modified})]. This
requires the knowledge of the asymptotic properties of the solution $%
\widehat{\Phi }_{x_{o}}(x)$ in the limit $x_{o}\rightarrow 0^{+}$.
The following Lemma will be invoked \cite{Tessarotto2006}:

\subsection{LEMMA - Asymptotic properties of $\widehat{\Phi}_{x_{o}}(x)$}

\emph{For any strong solution of the modified DSP, }$\widehat{\Phi }%
_{x_{o}}(x)$ \emph{obtained letting} $x_{p}=0,$\emph{\ the limit function }$%
\lim_{x_{o}\rightarrow 0^{+}}\widehat{\Phi }_{x_{o}}(x)$\emph{\
has the following properties:}

\emph{1) There results:}%
\begin{equation}
\lim_{x_{o}\rightarrow0^{+}}\widehat{\Phi}_{x_{o}}(x_{o})=+\infty;
\label{limit-1}
\end{equation}

\emph{2) for any }$x>0$ \emph{the integral limit (\ref{Eq.3-3}) is
satisfied
by }$\widehat{\Phi}_{x_{o}}(x).$\emph{This implies that the limit function }$%
\widehat{\Phi}(x)=\lim_{x_{o}\rightarrow0^{+}}\widehat{\Phi}_{x_{o}}(x)$
\emph{results such for any }$x>0,x\in_{\left\{ 0\right\} }$%
\begin{equation}
\widehat{\Phi}(x)=\lim_{x_{o}\rightarrow0^{+}}\widehat{\Phi}_{x_{o}}(x)=0.
\label{limit-3c}
\end{equation}

\emph{3) the following limit is satisfied by the boundary value }$\widehat{%
\Phi}_{x_{o}}(x_{o})$%
\begin{equation}
\lim_{x_{o}\rightarrow0^{+}}x_{o}\widehat{\Phi}_{x_{o}}(x_{o})=0.
\label{limit-4a}
\end{equation}

\emph{4) the limit value of the effective dimensionless charge }$%
c(x_{o},\Gamma )$\emph{\ for }$x_{o}\rightarrow 0^{+}$\emph{,
obtained
keeping }$\Gamma $ \emph{finite, is}%
\begin{equation}
\lim_{x_{o}\rightarrow 0^{+}}c(x_{o},\Gamma )=0.  \label{limit-5}
\end{equation}

PROOF

In fact, as a consequence of the integral equation
(\ref{constraint for Fi}) and the continuous dependence of
$\widehat{\Phi }_{x_{o}}(x)$ on the initial
data, it follows%
\begin{equation}
\lim_{x_{o}\rightarrow 0^{+}}x_{o}\widehat{\Phi
}_{x_{o}}(x_{o})=\beta -\lim_{x_{o}\rightarrow
0^{+}}x_{o}\int_{x_{o}}^{\infty }dx^{\prime
}x^{\prime }\sinh \widehat{\Phi }_{x_{o}}(x^{\prime })\widehat{\Theta }%
(x^{\prime }-x_{o}),  \label{limit-4b}
\end{equation}%
which implies
\begin{align}
& \left. \lim_{x_{o}\rightarrow 0^{+}}\widehat{\Phi
}_{x_{o}}(x_{o})=\infty
,\right. \\
& \left. \lim_{x_{o}\rightarrow 0^{+}}\int_{x_{o}}^{\infty
}dx^{\prime }x^{\prime }\sinh \widehat{\Phi
}_{x_{o}}(x)\widehat{\Theta }(x^{\prime }-x_{o})=\infty ,\right.
\end{align}%
i.e., the limit function $\lim_{x_{o}\rightarrow 0^{+}}\widehat{\Phi }%
_{x_{o}}(x)$ diverges in $x=x_{o}.$ Therefore, due to the continuity of \ $%
\widehat{\Phi }_{x_{o}}(x)$ with respect to $x\in \left[ 0,\infty
\right[ $ it follows that infinitesimally close to $x,x_{o}=0,$
and when $x,x_{o}$ are infinitesimal of the same order,
$\widehat{\Phi }_{x_{o}}(x)$ must diverge logarithmically as
\begin{equation}
\widehat{\Phi }_{x_{o}}(x)\sim \ln \left\{ \frac{1}{x^{3}}\right\}
. \label{divergency}
\end{equation}%
Let us now consider the implications of the integral equation
(\ref{integral
form - DSP modified}) for the limit function \ $\lim_{x_{o}\rightarrow 0^{+}}%
\widehat{\Phi }_{x_{o}}(x)$ for arbitrary $x\in _{\left\{
x_{o}\right\} }.$
There follows%
\begin{align}
\lim_{x_{o}\rightarrow 0^{+}}\widehat{\Phi }_{x_{o}}(x)& \mathbf{=}\frac{%
\beta }{x}-\frac{1}{x}\lim_{x_{o}\rightarrow
0^{+}}\int_{x_{o}}^{x}dx^{\prime }x^{\prime 2}\sinh \widehat{\Phi }%
_{x_{o}}(x^{\prime })\widehat{\Theta }(x^{\prime }-x_{o})- \\
& -\int_{x}^{\infty }dx^{\prime }x^{\prime }\lim_{x_{o}\rightarrow
0^{+}}\sinh \widehat{\Phi }_{x_{o}}(x^{\prime })\widehat{\Theta
}(x^{\prime }-x_{o}),  \notag
\end{align}%
where, due to the asymptotic estimate (\ref{divergency}), the
second term on the r.h.s. necessarily diverges
\begin{equation}
\lim_{x_{o}\rightarrow 0^{+}}\frac{1}{x}\int_{x_{o}}^{x}dx^{\prime
}x^{\prime 2}\sinh \widehat{\Phi }_{x_{o}}(x^{\prime })\widehat{\Theta }%
(x^{\prime }-x_{o})=\infty
\end{equation}%
unless there results for any $x\neq x_{o},$ $x\in _{\left\{
x_{o}\right\} }$
\begin{equation}
\lim_{x_{o}\rightarrow 0^{+}}\widehat{\Phi }_{x_{o}}(x)=0.
\label{limit-4 A}
\end{equation}%
As a consequence of Eq.(\ref{limit-4 A}), from the integral equation (\ref%
{integral form - DSP modified}) it follows necessarily that for all $x>0$:%
\begin{equation}
\lim_{x_{o}\rightarrow 0^{+}}\frac{\beta \widehat{\Theta }(x-x_{o})}{x}%
=\lim_{x_{o}\rightarrow
0^{+}}\frac{1}{x}\int_{x_{o}}^{x}dx^{\prime
}x^{\prime 2}\sinh \widehat{\Phi }_{x_{o}}(x^{\prime })\widehat{\Theta }%
(x^{\prime }-x_{o}),
\end{equation}%
which proves the limit (\ref{Eq.3-3}). As a consequence it must
result necessarily that the limit $\lim_{x_{o}\rightarrow
0^{+}}\sinh \widehat{\Phi }_{x_{o}}(x)$ is a Dirac delta. The
limit (\ref{limit-4a}) follows
immediately from the boundary condition (\ref{constraint for Fi}), while Eq.(%
\ref{limit-4 A}) implies manifestly the limit (\ref{limit-5}).

As an immediate consequence of ther Lemma it follows that the DSP
equation obtained letting $x_{o}=0$ in Eq.(\ref{integral form -
DSP modified}) does not admit classical solutions.

\subsection{THEOREM - Non-existence of classical solutions of DSP}

\emph{In the functional class }$\widehat{C}^{(2)}(_{\left\{ 0\right\} })$%
\emph{\ the DSP problem defined by Eqs.((\ref{integral form - DSP modified}%
)and the boundary conditions indicated above [Eqs.(BC-1 b)(BC-2
b)] does not admit strong solutions.}

PROOF

In fact, first, we notice that the limit function%
\begin{equation}
\lim_{x_{o}\rightarrow0^{+}}\widehat{\Phi}_{x_{o}}(x)\equiv\widehat{\Phi}(x),
\end{equation}
is manifestly a solution of the DSP equation which satisfies the
required boundary conditions (\ref{BC-1 b},ref{BC-2 b}). On the
other hand, due to the Lemma, this solution is discontinuous in
$x=x_{o}$ and results a distribution. Hence it is not a strong
(classical) solution of the modified DSP problem.

The basic implication of the Lemma and the theorem is that the DSP
equation, provided by the DH model, must be regarded as physically
unacceptable, since it does not admit strong solutions. In this
regard it should be noted that, as a basic principle, physically
acceptable of solutions of ordinary (or partial) differential
equations characterizing the classical theory of fields must be
suitably smooth strong solutions. The modified Debye screening
problem here defined, instead, exhibits smooth strong solutions
and therefore appears, from this viewpoint, consistent.

\section{Conclusions}

The essential implication of the present result is that the
customary mathematical model used for the investigation of the
Debye screening problem is incorrect. The correct mathematical
model requires, in fact, the treatment of the local plasma sheath,
for which a simple model is provided by Eq.(\ref{integral form -
DSP modified}). Important physical consequences follow. These
concern the correct estimate of the Debye screening effect, which
occurs close to the local plasma sheath and, particularly, for
highly-charged test particles immersed in strongly-coupled
plasmas. In fact, the modified DSP can be used to obtain
asymptotic estimates for the effective dimensionless charge
$c(x_{o},\Gamma )$ carried by the DH potential in strongly-coupled
plasmas \cite{Tessarotto2006b}. The resulting charge screening
effect appears produced by non-linear effects in the Poisson
equation. As a consequence, outside the Debye sphere (i.e., for
$x>1) $ the DH potential generated by highly charged test
particles in strongly-coupled plasmas results strongly reduced
with respect to the theoretical value observed in the
corresponding weakly-coupled systems.

%% BACKMATTER
%%%%%%%%%%%%%%%%%%%%%%%%%%%%%%%%%%%%%%%%%%%%%%%%

%% If the bibliography is produced without BibTeX comment out the

\section*{Acknowledgments}
The support of the ICTP (International Center of Theoretical
Physics, Trieste, Italy) through the ICTP/TRIL Program is
acknowledged. Research developed in the framework of MIUR
(Ministero Universit\'a e Ricerca Scientifica, Italy) PRIN Project
\textit{\ Fundamentals of kinetic theory and applications to fluid
dynamics, magnetofluiddynamics and quantum mechanics}, partially
supported (P.N.) by CMFD Consortium (Consorzio di
Magnetofluidodinamica, Trieste, Italy).

\pagebreak

\begin{center}
\textbf{Figure captions}
\end{center}

%\newline
\noindent \textbf{Figure 1 -} Comparison between $f_{c}(x),c(x_{o},\Gamma )$
and $c^{(a)(}(x_{o},\Gamma ).$ \ The data are normalized with respect to $%
\beta ,$ the normalized charge of the isolated test particle. The figure
concerns the case with $\beta =1$ and $x_{o}=0.05,$ yielding $\Gamma =20.$
The horizontal straight line represents the asymptotic estimate $%
c^{(a)(}(x_{o},\Gamma ),$ while the curve below it is the graph of $%
f_{c}(x). $ It follows $f_{c}(x_{o})\cong $ $0.564,$ while the asymptotic
value $c(x_{o},\Gamma )$ $\cong 0.493$ is reached approximately at $x\approx
0.3$, and the upper bound for the normalized effective charge is $%
c^{(a)(}(x_{o},\Gamma )\cong 0.589$. \newline

\noindent\textbf{Figure 2 -} Comparison between $f_{c}(x),c(x_{o},\Gamma )$
and $c^{(a)(}(x_{o},\Gamma )$ for $\beta =5$ and $x_{o}=0.2$ (with $\Gamma
=25$)$.$ In this case $f_{c}(x_{o})\cong $ $0.369,$ while the asymptotic
value $c(x_{o},\Gamma )$ $\cong 0.28$ is reached approximately at $x\approx
0.4$, and $c^{(a)(}(x_{o},\Gamma )\cong 0.38$. \newline

\noindent\textbf{Figure 3 -} Comparison between $f_{c}(x),c(x_{o},\Gamma )$
and $c^{(a)(}(x_{o},\Gamma )$ for $\beta =10$ and $x_{o}=0.3$ (with $\Gamma
\cong33$)$.$ In this case it is found $f_{c}(x_{o})\cong $ $0.273,$ while
the asymptotic value $c(x_{o},\Gamma )$ $\cong 0.188$ is reached
approximately at $x\approx 0.5$, and $c^{(a)(}(x_{o},\Gamma )\cong 0.303.$

\pagebreak
\begin{figure}[tbp]
\begin{center}
\includegraphics[width=.5\textwidth]{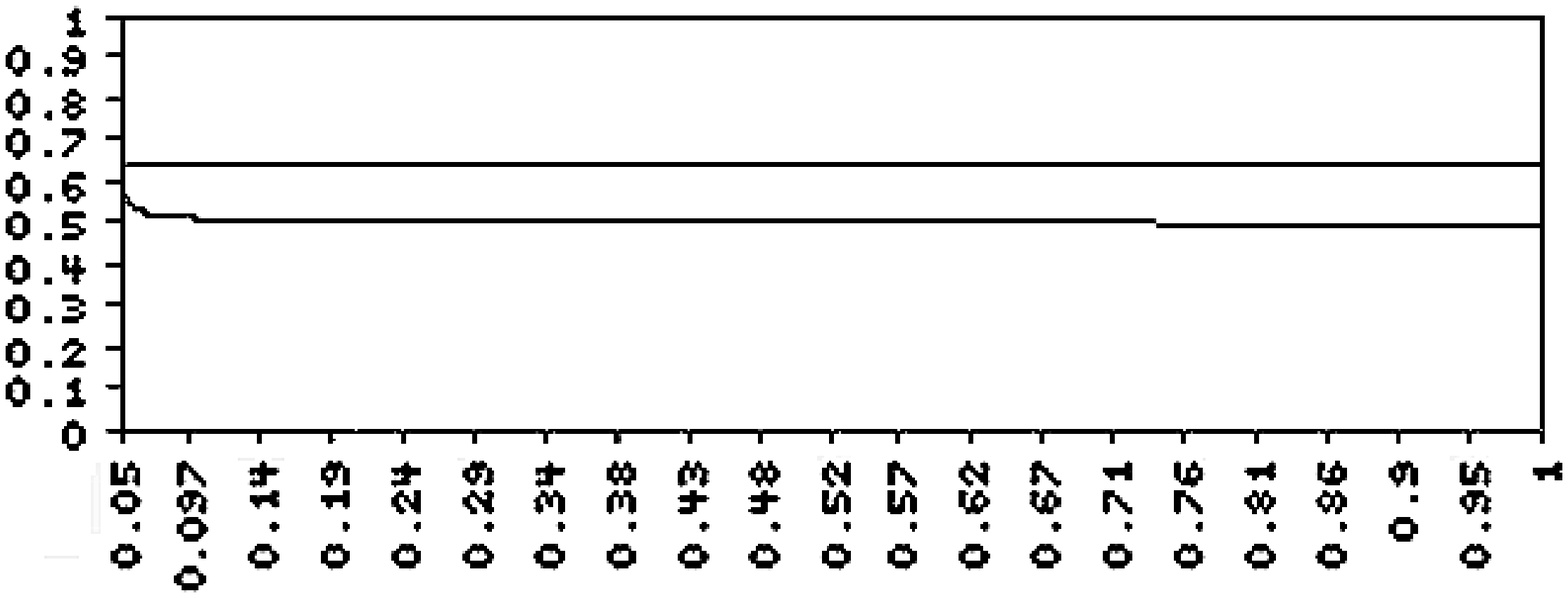}
\end{center}
\par
\vspace{-10pt}
\caption{}
\label{fig:1}
\end{figure}

\begin{figure}[tbp]
\begin{center}
\includegraphics[width=.5\textwidth]{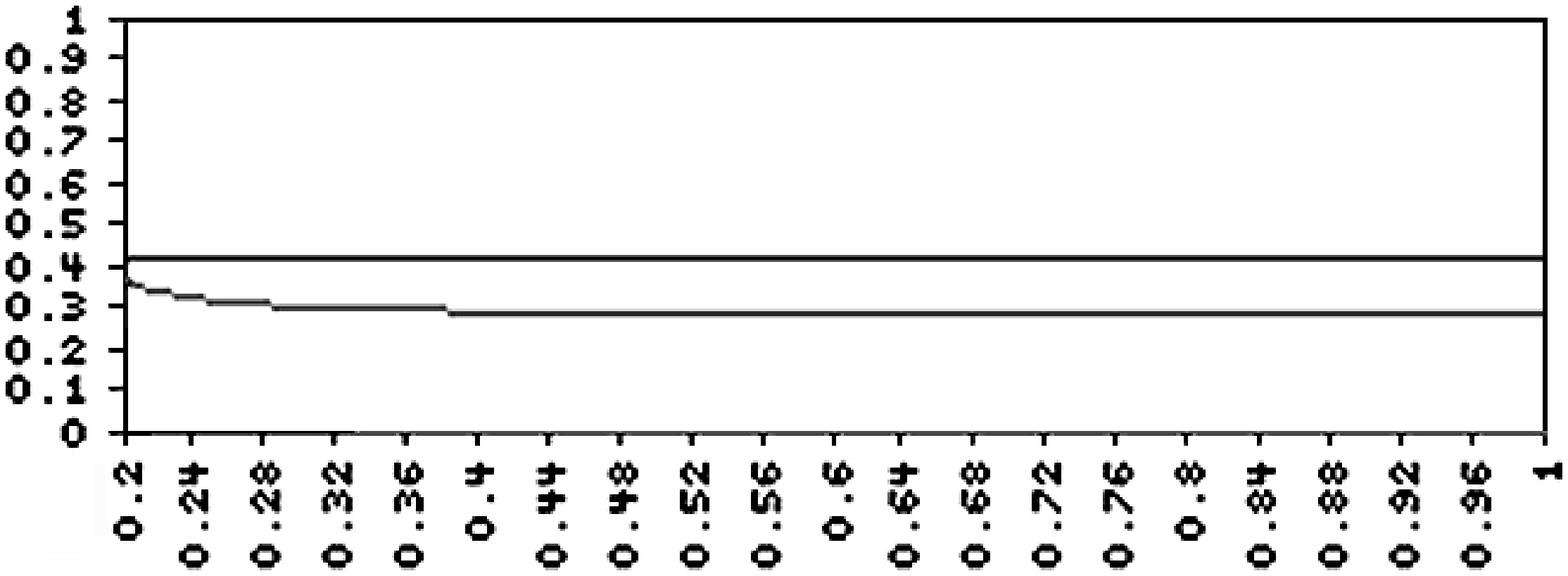}
\end{center}
\par
\vspace{-10pt}
\caption{}
\label{fig:2}
\end{figure}

\begin{figure}[tbp]
\begin{center}
\includegraphics[width=.5\textwidth]{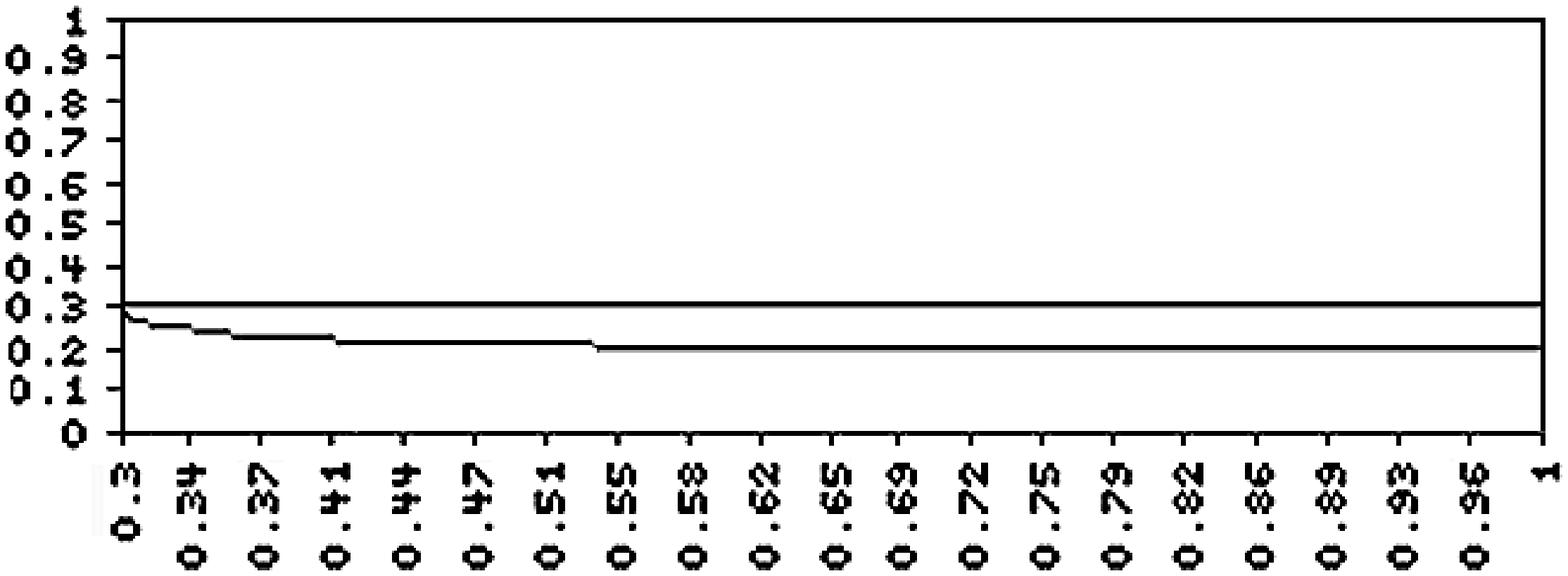}
\end{center}
\par
\vspace{-10pt}
\caption{}
\label{fig:3}
\end{figure}


\begin{thebibliography}{99}
\bibitem{Debye 1923} P. Debye and E. H\"{u}ckel, Phys. Zeit. \textbf{24}(9),
186 (1923); see also \textit{The collected papers of Peter J.W.
Debye }(Ox Bow Press, Conn., USA,1988), p.217.

\bibitem{Robbins-1988} M.O.Robbins, K.Kremer, G.S. Grest, J.Chem. Phys.
\textbf{88}, 3286 (1988).

\bibitem{Dupont 1996} G. Dupont, Mol. Phys.\textbf{79}, 453 (1996).

\bibitem{Allayrov 1998} E.Allayrov, H.L\"{o}wen, S.Trigger, Phys. Rev.E
\textbf{57}, 5518 (1998).

\bibitem{Bystrenko 1999} O. Bysternko and A. Zagorodny, Phys.Lett. \textbf{A
255}, 325 (1999).

\bibitem{Tsytovich 2001} P. Ricci, G. Lapenta, U de Angelis and V.N.
Tsytovich, Phy.Plasma \textbf{8}, 769 (2001).

\bibitem{Bystrenko 2003} O. Bysternko, T. Bysternko and A. Zagorodny,
Cond.Matter Physics \textbf{6 }(3), 425 (2003).

\bibitem{Tessarotto2006} D. Sarmah, M. Tessarotto and M. Salimullah, Phys.
Plasmas, \textbf{13}, 032102 (2006).

\bibitem{Tessarotto2006b} D. Sarmah, M. Tessarotto and M. Salimullah,
\textit{Non-linear charge reduction effect in strongly-coupled
plasmas}, http://www.arxiv.org/physics/0601077; in press Phys.
Scripta (2006).

\bibitem{Perez 1998} R.J. Perez and P. Martin, Astr.Space Sc.\textbf{\ 256},
263 (1998).

\bibitem{Chang Lin 2000} Chang Lin, Jin-bao Zhao and Xiu-Iian Zhang, Physica
Scripta \textbf{62}, 405 (2000).

\bibitem{Martin 1994} P. Martin and G.A. Baker Jr., J.Math.Phys. \textbf{32}%
, 1490 (1991).

\bibitem{Lampert 1980} M.A. Lampert and R.S. Crandall, Phys.Rev. A \textbf{21%
}, 362 (1980).

\bibitem{Lo Surdo 1994} C. Lo Surdo, A. Nocentini, Proceedings of the 1994
International Conference of Plasma Physics, edited by. P.H.
Sakanaka, E. Del Bosco and M.V. Alves (INPE, Sao Jose' dos Campos,
Brazil, 1994), p.163.

\bibitem{Garrett 1988} A.J. Garrett, Phys. Rev.A \textbf{37}, 4354 (1988).
\bibitem{Bernstein1959} I.B. Bernstein, \ I.N. Rabinowitz, Phys.Fluids
\textbf{2}, 112 (1959).

\bibitem{Alpert1965} Ya.L. Al'pert, A.V.Gurevich and L.P. Pitaevskii,
\textit{Space Physics and Artificial Satellites} (Plenum Press,
N.Y. 1965).

\bibitem{Allen1992} J.E.Allen, Physica Scripta \textbf{45}, 497 (1992).

\bibitem{Allen1999} J.E.Allen, B.M.Annaratone and U.de Angelis, J.Plasma
Phys. \textbf{63 (4)}, 299 (1999).

\bibitem{Tsytovich2005} V.N. Tsytovich and N.G.Gusein-zade, Plasma Phys.
Rep. \textbf{31} (10), 824 (2005).

\bibitem{Tessarotto 1992} A. Gregoratto, C. Lo Surdo, M. Tessarotto and R.
Zorat, Proceedings of the 2nd Symposium on Plasma Dynamics: Theory
and Applications, edited by M. Tessarotto (C.D.C., Udine, Italy,
1992), p.39.
%\bibliographystyle{aipproc}
%\bibliography{sample}
\end{thebibliography}
\end{document}